# Room Temperature Polariton Lasing in All-Inorganic Perovskites

Rui Su[1], Carole Diederichs[2,3], Jun Wang[4], Timothy C.H. Liew[1], Jiaxin Zhao[1], Sheng Liu[1], Weigao Xu[1], Zhanghai Chen[4], Qihua Xiong[1,2,5,*]

[1]Division of Physics and Applied Physics, School of Physical and Mathematical Sciences, Nanyang Technological University, 637371, Singapore.

[2]MajuLab, CNRS-UNS-NUS-NTU International Joint Research Unit, UMI 3654, Singapore

[3]Laboratoire Pierre Aigrain, Ecole normale supérieure, PSL Research University, CNRS, Université Pierre et Marie Curie, Sorbonne Universités, Université Paris Diderot, Sorbonne Paris-Cité, 24 rue Lhomond, 75231 Paris Cedex 05, France

[4]State Key Laboratory of Surface Physics, Department of Physics, Fudan University, Shanghai 200433, People's Republic of China

[5]NOVITAS, Nanoelectronics Center of Excellence, School of Electrical and Electronic Engineering, Nanyang Technological University, 639798, Singapore.

*To whom correspondence should be addressed. E-mail address: Qihua@ntu.edu.sg

**Polariton lasing is the coherent emission arising from a macroscopic polariton condensate first proposed in 1996[1]. Over the past two decades, polariton lasing has been demonstrated in a few inorganic[2-9] and organic semiconductors[10-13] in both low and room temperatures. Polariton lasing in inorganic materials significantly relies on sophisticated epitaxial growth of crystalline gain medium layers sandwiched by two distributed Bragg reflectors in which combating the built-in strain and mismatched thermal properties is nontrivial. On the other hand, organic active media usually suffer from large threshold density and weak nonlinearity due to the Frenkel exciton nature[11]. Further development of polariton lasing towards technologically significant applications demand more accessible materials, ease of device fabrication and broadly tunable emission at room temperature. Herein, we report the experimental realization of room-temperature polariton lasing based on an epitaxy-free all-inorganic cesium lead chloride perovskite microcavity. Polariton lasing is unambiguously evidenced by a superlinear power dependence, macroscopic ground state occupation, blueshift of ground state emission, narrowing of the linewidth and the build-up of long-range spatial coherence. Our work suggests considerable promise of lead halide perovskites towards large-area, low-cost, high performance room temperature polariton devices and coherent light sources extending from the ultraviolet to near infrared range.**

Cavity exciton polaritons are bosonic quasiparticles resulted from the strong coupling between excitons and confined cavity photon modes[14,15]. The half-light, half-matter nature of exciton polaritons gives rise to their extremely light effective mass, typically only $10^{-4}$~$10^{-5}$ times the mass of an electron, which is regarded as a crucial feature for the realization of Bose–Einstein condensation (BEC) of polaritons at high temperatures[16]. BEC of polaritons is well-known to occur when the ground state is macroscopically occupied upon reaching a critical threshold and the distribution of occupancy follows a Bose–Einstein distribution. Usually, influenced by the short lifetime of polaritons, the distribution is altered by non-equilibrium effects and uncondensed excitons, resulting in low threshold coherent emission or the so-called polariton lasing[17]. Unlike conventional photonic lasing, population inversion is no longer a prerequisite for achieving polariton lasing, attributed to the mechanism of bosonic final-state stimulation. This distinct feature gives rise to significant low threshold of polariton lasing, compared to vertical cavity surface emitting lasing (VCSEL) which shares precisely the same cavity structure as polariton laser.

Polariton condensation and polariton lasing in solid state systems were first demonstrated in CdTe[2] and GaAs[4] quantum well microcavities. However, the operation of polariton condensation based on such semiconductors with Wannier-Mott excitons, such as CdTe[2], GaAs[4], and more recently InP[5], is limited at cryogenic temperatures due to the small exciton binding energy. On the contrary, robust excitons, as well as large oscillator strengths, promote further room temperature realizations of polariton condensation and polariton lasing in ZnO[6] and GaN[3] bulk microcavities. Nonetheless, inorganic planar microcavities usually require sophisticated epitaxial techniques to ensure the high quality of the microcavity as well as the optical gain medium, in which one has to combat the challenges of built-in strain and mismatch of thermal expansion coefficients. In contrast, organic materials exhibit Frenkel excitons with much larger exciton binding energy and ease of fabrication, provide alternative systems for achieving room temperature polariton emission[18-20] and polariton condensation, *e.g.* crystalline anthracene[10], amorphous 2,7-bis[9,9-di(4-methylphenyl)-fluoren-2-yl]-9,9-di(4-methylphenyl) fluorine (TDAF) molecules[11], ladder-type conjugated MeLPPP polymer[12] and enhanced green fluorescent protein[13]. Nevertheless, Coulomb interaction[21], which contributes primarily to polariton-polariton interactions, is significantly weaker in organic materials due to the Frenkel exciton nature, leading to inefficient polariton relaxation to ground state[22]. Therefore, a much higher threshold and weaker nonlinearity are often obtained in organic microcavities[23], compared with inorganic microcavities. In addition, there have also been tremendous and long-lasting interests in the past decade in using hybrid organic-inorganic perovskites for polariton emission[24-28] and in investigating polariton condensation and polariton lasing operated at room temperature in such materials, as they combine the advantages of both inorganic and organic materials, such as ease of synthesis, large exciton binding energies and excellent optical properties determined by the inorganic component. However, up to now, only room temperature strong coupling regime was observed[24-28] without any successful realization of polariton condensation, which is probably caused by the insufficient crystalline quality induced by solution chemistry. Recently, we have shown that all-inorganic lead halide perovskites grown by an epitaxy-free vapor phase technique exhibit excellent optical gain properties with large exciton binding energy and oscillator strength, tunable emission band from ultraviolet to near infrared, and better optical stability under high laser flux illumination compared with hybrid perovskites[29]. Therefore, we believe that the all-inorganic perovskites are ideal alternatives for oxide and nitride inorganic materials to accomplish scalable and high performance polariton lasing operated at room temperature.

Here, the studied microcavity structure (schematically shown in Figure 1a) consists of a 373 nm thick lead chloride perovskite ($CsPbCl_3$) nanoplatelet embedded in bottom and top DBRs made of 13 and 7 $HfO_2/SiO_2$ pairs, respectively. The perovskite nanoplatelet was grown on a muscovite mica substrate by a chemical vapor deposition method[30,31], and transferred to the bottom DBR by a dry transfer process. The full perovskite microcavity fabrication is completed following the sequential deposition of a top DBR by e-beam evaporation. We stress that highly crystalline $CsPbCl_3$ platelet can also be grown on bottom DBR structures directly, suggesting the compatibility with device fabrication (See Methods and Supplementary Section 1). No additional spacer is needed as the thickness of the perovskite nanoplatelet can be controlled thick enough to support Fabry-Pérot oscillations. Figure 1b schematically displays the angle-resolved microphotoluminescence Fourier imaging setup. Figure 1c shows a typical rectangular-shaped $CsPbCl_3$ nanoplatelet with a width of 12 μm after cavity fabrication, along with the corresponding fluorescence image shown in Figure 1d. Figure 1e depicts the room temperature photoluminescence and absorption spectra of the $CsPbCl_3$ perovskite nanoplatelet on mica substrate. The absorption exhibits a strong, narrow inhomogeneously broadened excitonic absorption peak at ~ 3.045 eV, which strongly suggests the large exciton binding energy at room temperature. Previous experimental and theoretical studies have indeed shown that excitons in $CsPbCl_3$ perovskite possess a large binding energy of about 72 meV[29,32], even larger than in GaN and ZnO[33]. The photoluminescence emission of $CsPbCl_3$ perovskite is centered at ~2.99 eV with a FWHM of 80 meV. While after being embedded into a microcavity, the ground state emission of $CsPbCl_3$ perovskite microcavity shifts to ~ 2.9 eV with the FWHM narrowing to 9.7 meV, corresponding to a quality factor of Q ~ 300.

We investigated the polaritonic behaviors of the $CsPbCl_3$ perovskite microcavity by angle-resolved reflectivity and photoluminescence measurements (equivalently as a function of the in-plane wavevector $k_\parallel$) at room temperature (T = 300 K). Figure 2a and 2b show the $k$-space mappings of angle-resolved reflectivity and photoluminescence spectrum, respectively. The dispersion obtained in photoluminescence mapping agrees well with that of reflectivity measurements, except for small energy differences due to slightly different locations of the excitation spots. Particularly, we did not observe any uncoupled exciton peak located at ~ 3.045 eV. The upper polariton dispersion can hardly be identified in both reflectivity and photoluminescence mappings, which is a broadly met situation in microcavities with large Rabi splitting energy[33], due to absorption in the electron-hole continuum, thermal relaxation and the high reflectivity of the top DBR (Supplementary Section 2). However, the lower polariton dispersion features in both reflectivity and photoluminescence mappings demonstrate unambiguously the strong exciton-photon coupling regime: (i) the dispersion curvature tends to be smaller and smaller at large angles; (ii) an inflection point is distinguished near angle $\sim \pm 50°$, resulted from the onset of anticrossing between the perovskite exciton and the Fabry-Pérot cavity mode in strong coupling regime. It is noted that another parabolic-like dispersion in reflectivity mapping is attributed to the bare microcavity without any embedded perovskite (Supplementary Section 3), owing to the smaller size of the $CsPbCl_3$ perovskite nanoplatelet compared to the white light beam diameter. The dashed white lines in Figures 2a and 2b follow the theoretical fitting dispersion of the upper and lower polariton dispersions based on the coupled harmonic oscillator model[16]. The solid white lines display the dispersions of the uncoupled $CsPbCl_3$ perovskite exciton ($E_{ex} = 3.045\ eV$) and cavity photon modes obtained from the fitting. From the fitting of the data, we extract a Rabi splitting energy of $\sim 2\Omega = 265\ meV$ and a negative exciton-photon detuning of $\Delta = -25\ meV$, which corresponds to a photon fraction of 0.55 at the lower polariton branch minimum.

To reach the emission in the nonlinear regime, the system was pumped by off-resonant excitation centered at 375 nm with a pulse duration of 100 fs and a spot size of 25 μm. Figures 3a-3c display the contour maps of the angle-resolved photoluminescence at pumping fluence corresponding to a factor 0.8, 1.0 and 1.3 of the threshold fluence ($P_{th} = 12\ \mu J/cm^2$), respectively. At the low pump fluence of 0.8 $P_{th}$, the lower polariton dispersion exhibits a broad emission distribution at all angles. Upon reaching the threshold fluence, the polariton dispersion near $k_{\parallel} = 0$ exhibits a much stronger emission along with a sharp increase of intensity, indicating the onset of polariton condensation. Under even higher pump fluence of 1.3 $P_{th}$, the ground state near the minimum of lower polariton dispersion ($E_L(k_\parallel = 0) = 2.90\ eV$) becomes massively occupied, in stark contrast to the conventional photonic lasing which should massively occupy the minimum of the cavity mode dispersion ($E_C(k_\parallel = 0) = 3.02\ eV$). The enhanced macroscopic ground state occupation at the minimum of the polariton dispersion is one of the main features for polariton lasing.

Figure 4a shows the ground state emission spectra in logarithmic scale, measured at normal incidence ($k_\parallel = 0$) under different pump fluences. With the increase of pump fluence, the ground state emission witnesses a dramatic increase of intensity, along with a narrowing of the FWHM and a continuous blueshift in emission energy, suggesting a transition from a linear regime to nonlinear regime. To investigate this transition quantitatively, we plot the emission intensity at $k_\parallel = 0$ as a function of pump fluence in a log-log scale (red trace in Fig. 4b). A sharp increase, by three orders of magnitude of the ground state emission intensity is observed in the output-input characteristics above a threshold of $P_{th} = 12\ \mu J/cm^2$. Remarkably, despite the low Q of our microcavity, this threshold is at least five times lower than previous reports at room temperature in epitaxy-free microcavities[10-12]. Meanwhile, the FWHM of the ground state emission (blue trace in Fig. 4b) first exhibits a slight broadening below threshold, due to polariton-polariton interaction[34], then, it dramatically narrows from 10.3 meV to 3.4 meV by further increasing the pump fluence above threshold to 1.5 $P_{th}$, indicating a sharp increase of the temporal coherence. Beyond 1.5 $P_{th}$, the FWHM broadens slightly again, which could be attributed to decoherence induced by polariton self-interactions[34], as well as different localized condensate modes[2], such as in previous studies on inorganic microcavities. The threshold and linewidth features clearly suggest the occurrence of stimulated scattering into the polariton ground state at $k_\parallel = 0$. Another crucial evidence to prove the appearance of polariton lasing is the continuous blueshift of the polariton emission energy along with the increase of excitation density, which serves as a clear signature of the repulsive interactions between the polaritons. We plot the energy blueshift of the ground state emission at $k_\parallel = 0$ as a function of pump fluence as shown in Figure 4c. We observe a maximum blueshift of 10 meV of the polariton condensate, which is much smaller than the energy difference of 120 meV between the minimum of the lower polariton and uncoupled cavity mode dispersions. Two distinct slopes are also observed below and above threshold, and to gain more insight into this behavior, we developed a model by coupling the driven-dissipative Gross-Pitaevskii equation to a rate equation for a reservoir of excitons (Methods). The calculated blueshift, shown as a solid red line in Figure 4c, fits well with our experimental results. Below threshold, it is dominated by the interaction between polariton and the reservoir of excited states. While above the threshold, the polariton-polariton interaction becomes much more prominent, compared with the polariton-reservoir interactions. The difference of interaction strengths between these two mechanisms gives rise to these two distinct blueshift trends[11].

To further confirm the occurrence of polariton lasing unambiguously, we checked the long-range spatial coherence by interferometry measurements, as it is one of the important features of polariton condensation. For this purpose, the real space emission image of the polariton condensate is sent to a Michelson interferometer in which one of the arms is replaced by a retroreflector to invert the image in a centro-symmetrical configuration. The use of the retroreflector enables the measurement of the first-order spatial coherence $g^{(1)}(r, -r)$ based on the interference fringes contrast, or in other words the phase coherence between points localized at *r* and –*r* from the center of the polariton condensate. Figure 4d and 4e display the real space emission image of the polariton condensate and its inverted image, respectively. First of all, we observe several bright spots in the real space emission image, which are associated to the localization of the polariton condensate, possibly as a result of the photonic disorder inherited from the sample transfer process during the fabrication of the perovskite microcavity. Similar phenomena hve been widely observed in organic and inorganic polariton condensates before[2,10].

Furthermore, effects on the polariton distribution near $k_{\parallel}=0$ from disorder can also be identified from the lower polariton branch above the polariton lasing threshold of other typical samples on the same sample chip (Supplementary Section 4). We extended the Gross-Pitaevskii theory to account for spatial dynamics in the real space. Using the dispersion obtained from coupled oscillator fitting and including some disorder, the localization effect can be successfully reproduced (see Supplementary Section 5). Finally, Figure 4f presents the interference image resulting from the superposition of the two previous images shown in Figures 4d and 4e, where unambiguous interference fringes are readily identified along a distance as large as 15 μm. This latter result is a clear demonstration of the build-up of a long range spatial coherence associated to the formation of a polariton condensate in our all-inorganic perovskite system.

In conclusion, we have observed unambiguous evidence for polariton lasing in an all-inorganic $CsPbCl_3$ perovskite planar microcavity at room temperature. The successful realization of polariton lasing within a low Q microcavity, along with its ease of fabrication and the inorganic nature of its active medium, significantly alleviates the stringent requirements to approach the fundamental physics of BEC. Our findings open a new platform with compelling properties towards realization of large-area, low-cost and high-performance polariton devices and potentially towards an electrically pumped BEC coherent light source at room temperature.

## Methods:

**Microcavity fabrication.** The bottom DBR was first fabricated by an e-beam evaporator, consisting of 13 pairs of silicon dioxide (73 nm) and hafnium dioxide (54 nm) capped by silicon dioxide. It was then put as an in-situ substrate in a single zone tube furnace (Lindberg/Blue MTF55035C-1) to directly grow $CsPbCl_3$ perovskite crystals on top of the bottom DBR. The detailed procedure of the growth of the perovskite is the same as described in our previous report on mica substrate[29]. Thinner $CsPbCl_3$ perovskite platelets could also be transferred from growth mica substrate to bottom DBR substrates by a dry tape transfer process, while the Mica substrate was totally removed by exfoliation using a scotch tape. The bottom DBR with $CsPbCl_3$ perovskite platelet on top was then transferred into the e-beam evaporator again to complete the fabrication of a top DBR, which consists of 7 pairs of silicon dioxide (68 nm) and hafnium dioxide (50 nm). Both methods (in-situ growth or dry transfer) can work to produce high quality samples for cavity polariton studies (See Supplementary Section 1).

**Optical Spectroscopy Characterization.** The fluorescence image of the perovskite microcavity was obtained through an Olympus microscope where the microcavity is illuminated with an Olympus U-HGLGPS lamp after passing through a 355 nm bandpass filter. The absorption spectrum was measured using a PerkinElmer Lambda 950 UV-VIS-IR spectrometer. Steady-state photoluminescence measurement was conducted in a confocal spectrometer (Horiba Evolution) by using a Helium-Cadmium laser (325 nm) with a pump power of 10 μW. Angle-resolved photoluminescence and reflectivity spectroscopy was measured in a home built micro-photoluminescence setup within the Fourier imaging configuration. For a continuous wave laser of 325 nm pumping and white light illumination, the angle-resolved photoluminescence and reflectivity shown in Figure 2 were measured through a high numerical aperture 100× microscope objective (NA = 0.9), covering an angular range of ± 64.1°. For pulsed laser pumping, the angle-resolved photoluminescence shown in Figure 3 was measured through a long working distance 100× microscope objective (NA = 0.75), covering an angular range of ± 48.5°. The emission from the microcavity is collected through the same objective and sent to a 550 mm focal length spectrometer (Horiba iHR550) with a 600 lines/mm grating and a 256 × 1024 pixel liquid nitrogen cooled charge-coupled-device (CCD). The perovskite microcavity is pumped by off-resonant excitation centered at 375 nm with pulse duration of 100 fs and repetition rate of 1 KHz (Lightconversion optical parametric amplifier pumped by a Spectra Physics Spitfire Ace Ti:Sapphire regenerative amplifier). The excitation laser beam which is slightly elliptic was focused down to a ~25 μm spot on the sample.

**Nonequilibrium Condensation Model.** To describe the nonequilibrium condensation in our system we apply the driven-dissipative model of previous report[35], neglecting for simplicity spatial dependence:

$$i\hbar \frac{d\psi}{dt} = \left[ E_0 + \alpha |\psi|^2 + g_R n_R + \frac{i}{2}(\hbar r n_R - \Gamma) \right] \psi$$

$$\frac{dn_R}{dt} = P - \left( \Gamma_R + r|\psi|^2 \right) n_R$$

Here $\psi$ represents the coherent polariton field, coupled to an equation for the density of higher energy excitations $n_R$ (a reservoir). $E_0$ is the energy of the bare polariton resonance; $\alpha$ represents the strength of polariton-polariton interactions; $g_R$ the strength of interactions between polaritons and the high energy reservoir; $r$ is the condensation rate; $\Gamma$ is the polariton dissipation rate; $\Gamma_R$ is the reservoir decay rate; and $P$ is the pumping rate.

The equations are readily solved analytically for the steady state, where the condensate mean-field population $|\psi|^2$ is non-zero above the threshold pumping rate $P_0 = \frac{\Gamma \Gamma_R}{\hbar r}$. Above the threshold, the polariton energy shift with pumping rate is given by

$$\Delta E = \frac{\hbar P \alpha}{\Gamma} - \frac{\Gamma_R \alpha}{r} + \frac{g_R \Gamma}{\hbar r}.$$

And below the threshold, the polariton energy shift with pumping rate is given by

$$\Delta E = \frac{g_R P}{\Gamma_R},$$

this allows plotting the theoretical curve in **Figure 4c**, taking parameters $\frac{g_R \Gamma}{\hbar r} = 10.6 \ meV$ and $\frac{\Gamma_R \alpha}{r} = 0.86 \ meV$ as the relevant fitting parameters (it is also assumed that the theoretical pumping rate $P$ is linearly proportional to the experimental pump power). The slope of the energy shift with pumping rate has a discontinuity at the threshold, since below and above threshold the condensate and reservoir populations have different growth rates with pumping rate. In addition, the interaction energy associated with polariton-polariton interaction and polariton-reservoir interaction is generally different.

**Acknowledgements**

Q.X. acknowledges the support from the Singapore National Research Foundation through the NRF Investigatorship Award (NRF-NRFI2015-03), and the Singapore Ministry of Education via AcRF Tier 2 grant (MOE2015-T2-1-047). This work was also supported in part by a Competitive Research Program (NRF-CRP-6-2010-2) from the Singapore National Research Foundation. T.L. acknowledges support from the Singapore Ministry of Education via AcRF Tier 2 grant (2015-T2-1-055).

**Author contributions**

R.S., C.D. and Q.X. conceived the idea and designed the research; R.S., J.W. and Z.C. performed the DBR fabrication, sample growth and all the optical measurements; T.C.L. carried out the theoretical calculations. R.S., C.D., Z.C. and Q.X. analyzed the data, R.S., C.D., T.C.L and Q.X. wrote the manuscript. All authors discussed and commented on the manuscript.

**Additional information**

Supplementary information is available in the online version of the paper. Reprints and permissions information is available online at www.nature.com/reprints. Correspondence and requests for materials should be addressed to Q.X.

**Competing financial interests**

The authors declare no competing financial interests.

**Figures and Captions**

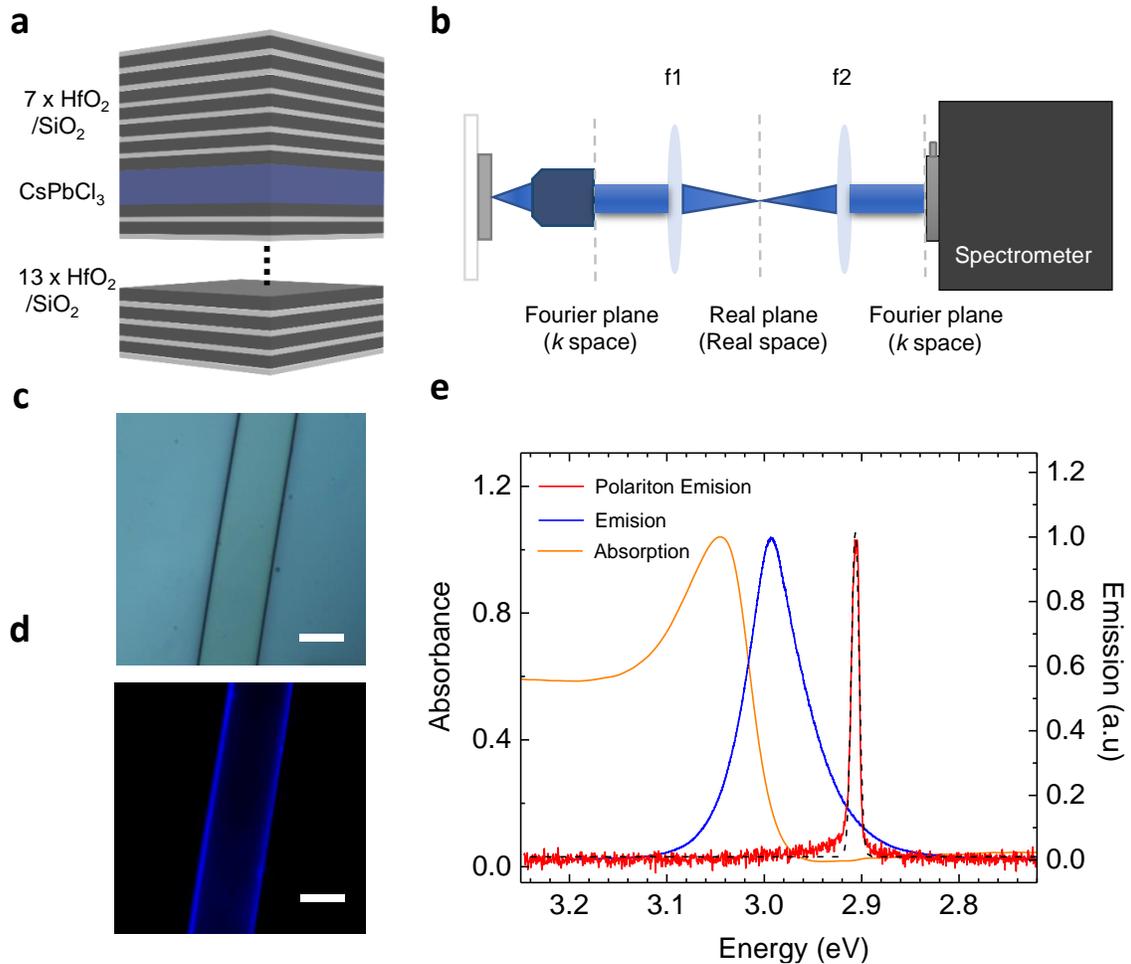

**Figure 1 Microcavity structure and emission. a**, Schematic of the perovskite microcavity, which consists of a $CsPbCl_3$ crystalline nanoplatelet embedded in a planar microcavity formed by a bottom and a top $HfO_2/SiO_2$ DBR. **b**, Schematic of the angle resolved microphotoluminescence setup with Fourier optics. Two lenses are shown as f1 and f2. **c**, Microscopy image of a rectangular-shaped perovskite microcavity after fabrication. Scale bar, 12 μm. **d**, Fluorescence microscopy image of as-fabricated perovskite microcavity, showing strong and uniform blue emission along the edges. Scale bar, 12 μm. **e**, Room temperature photoluminescence and absorption spectra of $CsPbCl_3$ perovskite nanoplatelet. Orange trace, the absorption spectrum of CsPbCl$_3$ perovskite nanoplatelet on mica substrate, showing a strong excitonic peak at ~ 3.045 eV. Blue trace, the photoluminescence emission spectrum of CsPbCl$_3$ perovskite nanoplatelet on mica substrate, showing an emission at ~ 2.99 eV with a FWHM of 80 meV. Red trace, the ground state emission with Gaussian fitting (black dash line) of CsPbCl$_3$ nanoplatelet after embedded into microcavity, showing an emission at ~ 2.90 eV with a FWHM of ~ 9.7 meV.

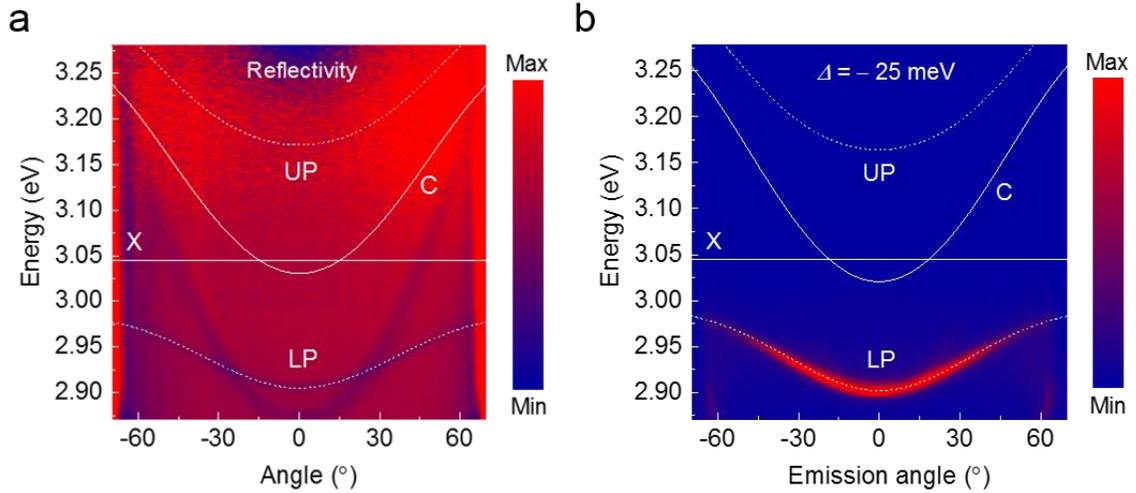

**Figure 2 Angle-resolved reflectivity and photoluminescence of $CsPbCl_3$ perovskite microcavity**. a, Angle-resolved reflectivity spectrum measured using a white light lamp. The dashed white lines show the theoretical fitting dispersion of the upper (UP) and lower (LP) polariton dispersions. The solid white lines display the dispersions of uncoupled $CsPbCl_3$ perovskite exciton (X) and cavity photon mode (C) obtained from a coupled harmonic oscillator model fitting. The parabolic-like dispersion is caused by the bare microcavity, without any embedded perovskite, as a result of the smaller sample size compared with the white light beam diameter. **b**, Angle-resolved photoluminescence spectrum of perovskite microcavity. The detuning $\Delta$, obtained from a coupled harmonic oscillator model fitting to the measured dispersion (dashed white lines), is indicated on the figure. Note the slight energy differences between reflectivity and photoluminescence spectrum due to slightly different spot positions.

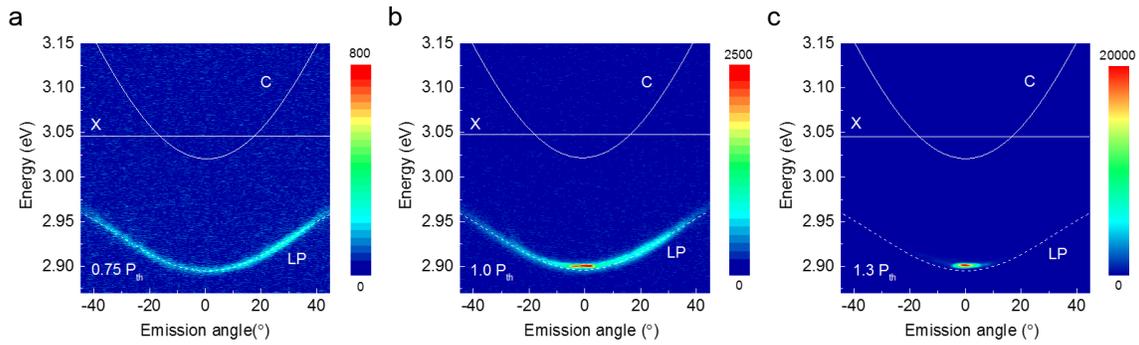

**Figure 3 Power dependent angle resolved photoluminescence spectrum. a**, Angle-resolved photoluminescence spectrum measured at 0.75 $P_{th}$. Polaritons show a broad emission distribution at all angles. **b**, Angle-resolved photoluminescence spectrum measured at 1.0 $P_{th}$. The ground state near $k_{\parallel} = 0$ exhibits a much stronger emission than other angles, indicating the onset of polariton lasing. **c**, Angle-resolved photoluminescence spectrum measured at 1.3 $P_{th}$. The ground state near $k_{\parallel} = 0$ is massively occupied, experiencing a sharp increase of intensity along with a blueshift of peak energy.

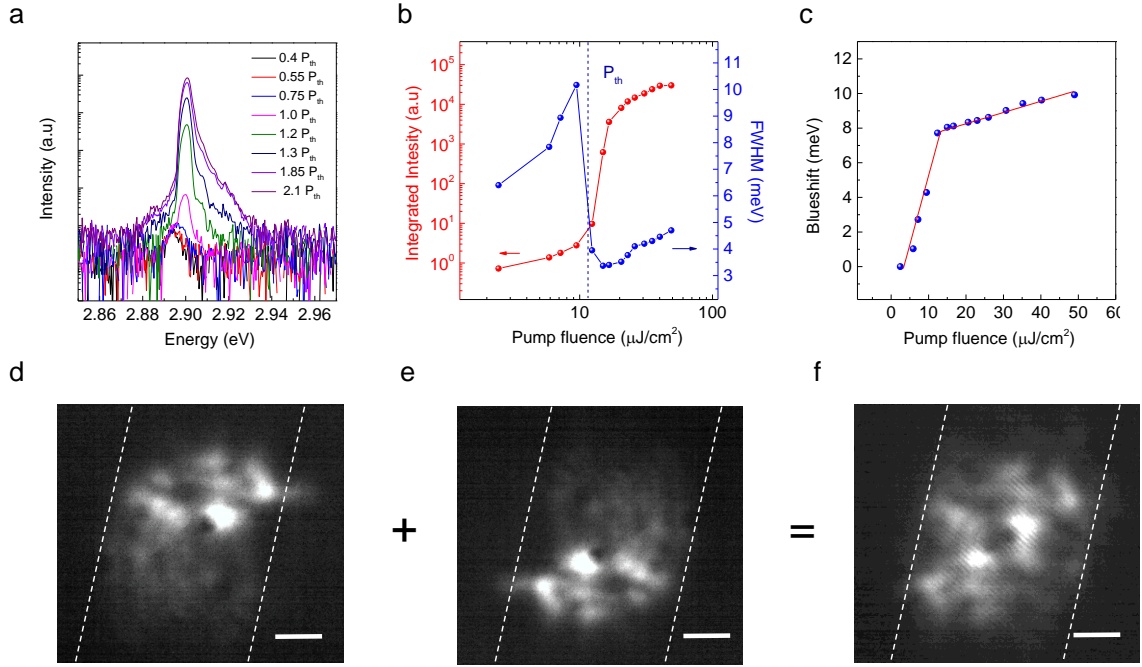

**Figure 4 Characterizations of CsPbCl$_3$ microcavity polariton lasing. a**, Ground state emission spectra at $k_\parallel = 0$ under different pumping fluences. A sharp increase of emission intensity occurs beyond 1.0 $P_{th}$, suggesting the transition to the nonlinear regime. The emission energy is observed to be blue-shifted with the increasing pump fluence. **b**, Ground state emission intensity at $k_\parallel = 0$ and full width at half maximum (FWHM) as a function of pump fluence. A linewidth narrowing occurs near the threshold of $P_{th}$ = 12 μJ/cm$^2$, along with a sharp increase of emission intensity. **c**, Energy blueshift with respect to the polariton emission energy at the lowest pump fluence as a function of pump fluence. The experimental blueshift of the ground state emission is shown as blue dots and the red solid line indicates the theoretical calculation of the blueshift. The blueshift trend below the threshold is attributed to polariton - reservoir interaction while the trend above the threshold corresponds to polariton-polariton interaction. **d**, Real space photoluminescence image measured above threshold from one arm of the Michelson interferometer. Dashed line, sample edges. Scale bar, 4 μm. **e**, Real space photoluminescence image obtained above threshold from the second arm by using a retroreflector to flip the image in a centrosymmetric way. Dashed line, sample edges. Scale bar, 4 μm. **f**, Interference pattern after superposition of the images above threshold from the two arms of the Michelson interferometer. Clear interferences are readily identified with a distance as large as 15 μm, demonstrating the build-up of a long range coherence. Dashed line, sample edges. Scale bar, 4 μm.